# Anomalous Electron Transport in Field-Effect Transistors with TiTe$_2$ Semimetal Thin-Film Channels

J. Khan[×], C.M. Nolen[+], D. Teweldebrhan[×], D. Wickramaratne, R. K. Lake and A.A. Balandin[∗]

Department of Electrical Engineering and Materials Science and Engineering Program

Bourns College of Engineering, University of California – Riverside

Riverside, California 92521 USA

[×]Present address: Intel Corporation, Hillsboro, Oregon, 97124 USA

[+]Present address: Intel Corporation, Chandler, Arizona, 85226 USA

---

[∗] Corresponding author; electronic address: balandin@ee.ucr.edu ; group web-site: http://ndl.ee.ucr.edu

**Abstract**

We report on "graphene-like" mechanical exfoliation of thin films of titanium ditelluride and investigation of their electronic properties. The exfoliated crystalline TiTe$_2$ films were used as the channel layers in the back-gated field-effect transistors fabricated with Ti/Al/Au metal contacts on SiO$_2$/Si substrates. The room-temperature current-voltage characteristics revealed strongly non-linear behavior with signatures of the source-drain threshold voltage similar to those observed in the charge-density-wave devices. The drain-current showed an unusual non-monotonic dependence on the gate bias characterized by the presence of multiple peaks. The obtained results can be potentially used for implementation of the non-Boolean logic gates.

Successful mechanical exfoliation of graphene [1] and investigation of its unique electrical [2-3] and thermal [4-7] properties motivated research of other quasi two-dimensional (2-D) materials. We have recently exfoliated quintuples – five atomic layers – of bismuth telluride ($Bi_2Te_3$) and related materials, which reveal thermoelectric and topological insulator properties [8-11]. It was shown that exfoliated $Bi_2Te_3$ films have electrical, thermal, and optical properties substantially different from those of bulk $Bi_2Te_3$ crystals [8-12]. The exfoliated thin films allow one to achieve strong quantum confinement of charge carriers and acoustic phonons owing to almost infinite potential barriers for electrons and phonons in such structures [8]. The molecular-beam-epitaxial films grown on the lattice matched substrates are normally characterized by small energy-band offsets [8-9]. Even if the exfoliated films are relatively thick – no quantum confinement effects – they represent interesting objects of investigation owing to their crystallinity and possibility of tuning the charge density via electrical gating.

The essential requirement for exfoliation of quasi-2-D films is a presence of the van-der-Waals gaps in the crystal structure. Here we report on the "graphene-like" exfoliation and investigation of physical properties of materials that have not been previously studied in such a context. We considered titanium dichalcogenides $TiCh_2$ (Ch=Te, Se, S), which are semimetals with the hexagonal crystal structure of the space group $D_{3d}^1$. The $TiCh_2$ trilayers are separated by the van der Waals gaps while inside each layer, Ti is surrounded by 6 chalcogen atoms in the octahedral configuration. These types of crystals – with atomic trilayers as their structural units – differs from both graphene – a single atomic layer – and $Bi_2Te_3$ containing a quintuple structure. Many transition-metal dichalcogenides [13-18] show phase transitions to a modulated crystal structure, referred to as a charge-density wave (CDW) [19-20]. We investigated properties of exfoliated $TiTe_2$ films with the help of the field-effect-transistor (FET) type devices. It is demonstrated that $TiTe_2$ films reveal unusual electronic properties, which can be potentially used for the information processing.

We exfoliated $TiTe_2$ films on $Si/SiO_2$ substrates following the standard "graphene-like" approach [1-2]. The thickness $H$ of the films ranged from a few trilayers (FTL) to $H\sim150$ nm. The thickness of the Te-Ti-Te atomic trilayer is $H$=0.271 nm while the lattice parameters are $a$ = 0.377 nm and $c$ = 0.6498 nm. The FTL films are more suitable for back-gating while the thicker films can be more readily incorporated into the FET structure. The atomic force microscopy (AFM) inspection (VEECO Dimension-5000) revealed that surface roughness of the films was ~1 nm. We have used micro-Raman spectroscopy to verify the crystallinity and quality of the flakes with the optimum thickness. Raman spectroscopy (Renishaw InVia) was performed in a backscattering configuration under $\lambda$ = 488-nm laser excitation. We used low excitation power on the sample surface (P<0.5 mW) to avoid local heating. The details of our Raman-spectroscopy procedures are reported elsewhere in the graphene-research context [21-23].

Figure 1 shows Raman spectra from TiTe$_2$ FTL film for the temperature *T* varying from ~165 K to ~425 K. One can clearly see two main peaks at ~120-124 cm$^{-1}$ and ~141-145 cm$^{-1}$. These peaks can be associated with the $E_g$ and $A_{1g}$ modes reported for bulk TiTe$_2$ and related materials [24]. It is interesting to note that we have not observed distinctive shoulders near ~160 cm$^{-1}$, which were previously attributed in TiCh$_2$ materials to poorly defined stoichiometry and defects [24]. The latter indicates the high purity of our samples. The crystallinity and elementary composition of the films were further confirmed by the energy dispersive spectroscopy (EDS) and the high-resolution scanning electron microscopy (SEM). The inset to Figure 1 is a representative SEM image (XL-30 FEG) of a TiTe$_2$ flake.

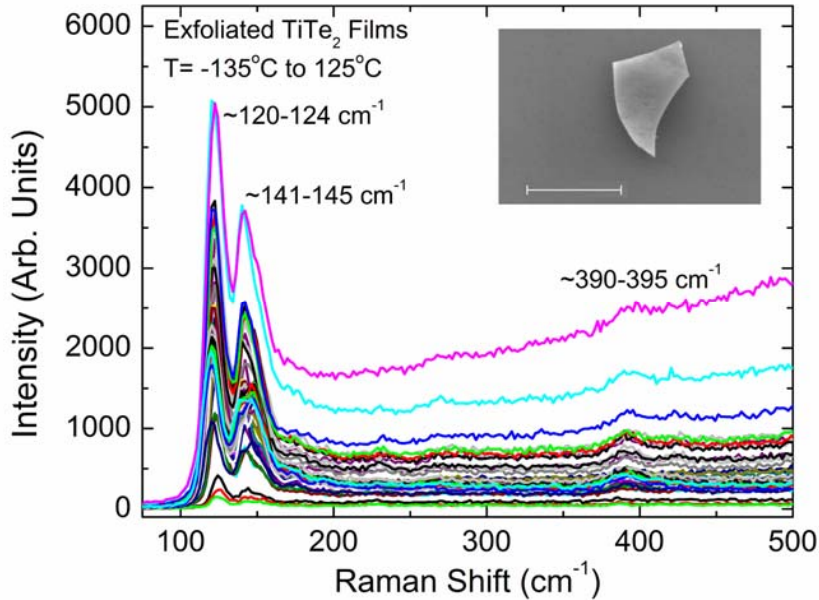

**Figure 1:** Raman spectra of the mechanically exfoliated TiTe$_2$ films at different temperatures. One can see two main $E_g$ and $A_{1g}$ phonon modes. The absence of the shoulder near ~160 cm$^{-1}$ suggests the high quality, stoichiometry and low defect concentration in the examined TiTe$_2$ films. The inset shows the SEM image of the "graphene-like" exfoliated TiTe$_2$ film.

The back-gated FET with TiTe$_2$-film channels were prepared by depositing the source, drain and back-gate metal contacts. The degenerately doped Si substrate acted as a back gate separated from the channel by the 300-nm-thick SiO$_2$ layer. The substrate was spin-coated (Headway SCE) with the copolymer (MMA) and polymethyl methacrylate (PMMA) for the top contact fabrication. The source and drain of the FET structure were defined by the electron beam lithography (Leo 1150). The metal layers Al/Ti/Au (8-nm/10-nm/100-nm) were sequentially deposited on the TiTe$_2$ in the defined contact patterns.

Since SiO$_2$ oxide and TiTe$_2$ channels were relatively thick we had to apply large gate biases. The room-temperature (RT) electrical resistance of the exfoliated films was ~$10^{-4}$ Ω–cm. The device structure is shown in the inset to Figure 2. In such a design the gate is global and affects not only the channel but the Al/Ti/Au – TiTe$_2$ contact as well. The current-voltage (I-V) characteristics of TiTe$_2$ devices are presented in Figure 2.

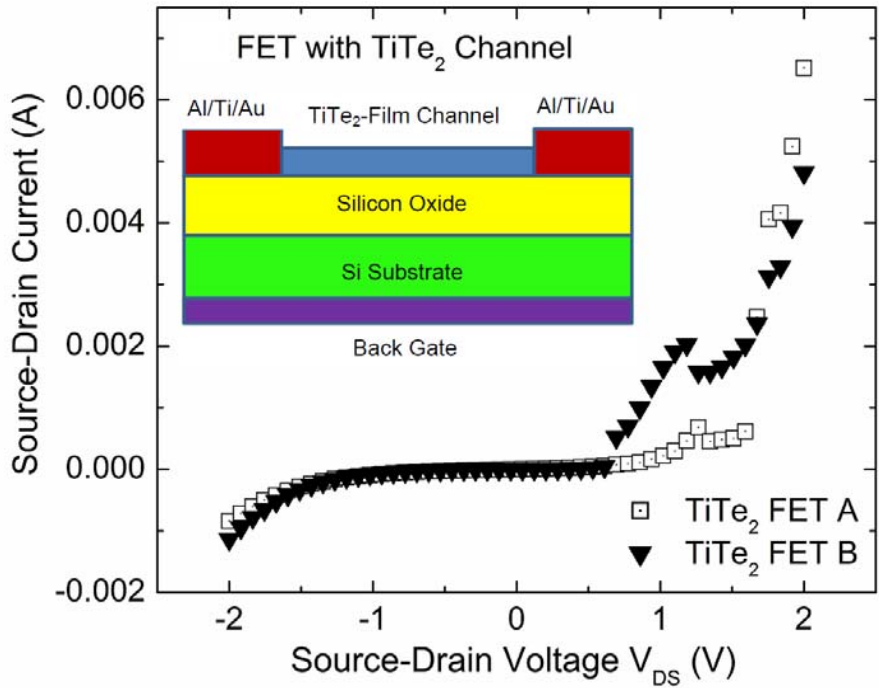

**Figure 2:** Drain current as the function of the source-drain voltage in back-gated FETs with TiTe$_2$ channel and Al/Ti/Au source and drain contacts. The measured I-V characteristics are strongly non-linear with a clear threshold voltage at $V_{DS}$=1.5 V ($V_{GS}$ = 50 V). The source-drain distance is 8 μm. The inset shows a schematic of the FET device with the TiTe$_2$ channel.

The curves marked A and B correspond to two different positions of the probe-station contacts on the source and drain electrodes. One can see that I-Vs are asymmetric and reveal a threshold voltage $V_{DS}$~1.5 V, where the conductance increases strongly. The measured I-Vs are reminiscent of those observed in CDW materials [19-20]. In CDW conductors, the threshold $V_{DS}$, indicates the electrical field $E_T$ at which the CDW depins and slides relative to the lattice [19]. The I-Vs are also similar to those resulting from tunneling contacts [25]. In this case, the Al/Ti/Au – TiTe$_2$ contact effects can be responsible for the observed threshold-type characteristics. It has been reported that oxidation of Ti layers used in the electrodes to CDW materials such as TaS$_3$ can lead to formation of a tunnel barrier resulting in non-linear I-Vs [25]. The existence of the tunnel barrier contacts is consistent with the sensitivity of the I-Vs to light exposure. Figure 3 presents the source-drain current as a function of $V_{DS}$ at RT. One can see that the exposure of the device to the direct light increases the low-bias current and reduces the non-linearity. The light is exciting carriers that can then either go over the tunnel barrier or more easily tunnel through the tunnel barrier with the net result being increased current and less non-linearity in the I-Vs.

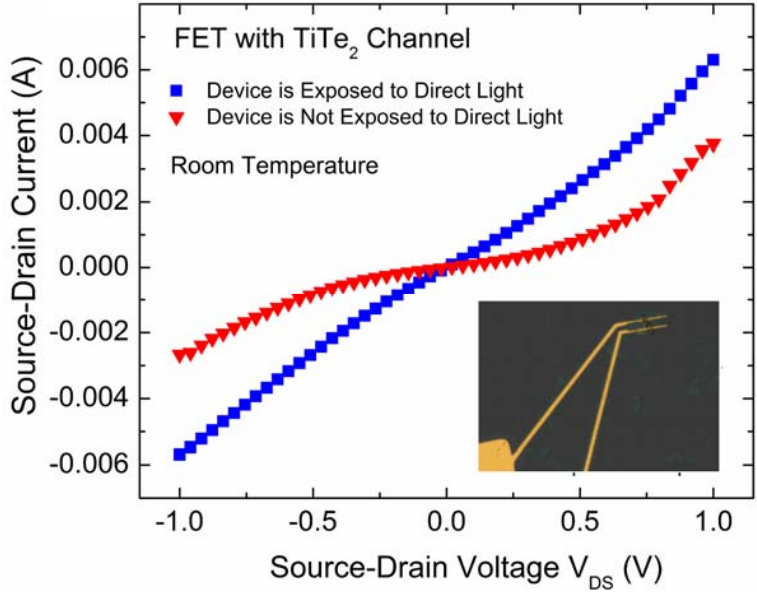

**Figure 3:** Drain current as the function of the source-drain voltage for the TiTe$_2$ FETs in the dark and under light illumination. The inset shows a SEM image of the tested device with Al/Ti/Au contacts and metal electrodes.

The back-gate $V_{GS}$ dependence of the drain-current $I_{DS}$ reveals unusual non-monotonic features – multiple and almost equidistant peaks (see Figure 4). Both the sensitivity of the channel current to $V_{GS}$ and the multiple, almost periodic peaks in the channel current as a function of $V_{GS}$ are difficult to explain. The periodicity suggests possible CDW effects, since a gate voltage can periodically modulate the phase of the CDW. We note that CDW effects in TiTe$_2$ are still the subject of debate [13, 15]. At $T_P$=150 K a phase transition was detected in the magnetic susceptibility measurements of the bulk TiTe$_2$, which was attributed to CDW [13]. It is not known how the Peierls transition temperature $T_P$ would depend on the thickness of TiTe$_2$ films when $H$ is in nanometer scale. It was shown that $T_P$ can increase in thin films owing to the formation of a single-domain CDW. The anomalous CDW depinning above $T_P$ was observed in thin crystals of NbSe$_3$ [26].

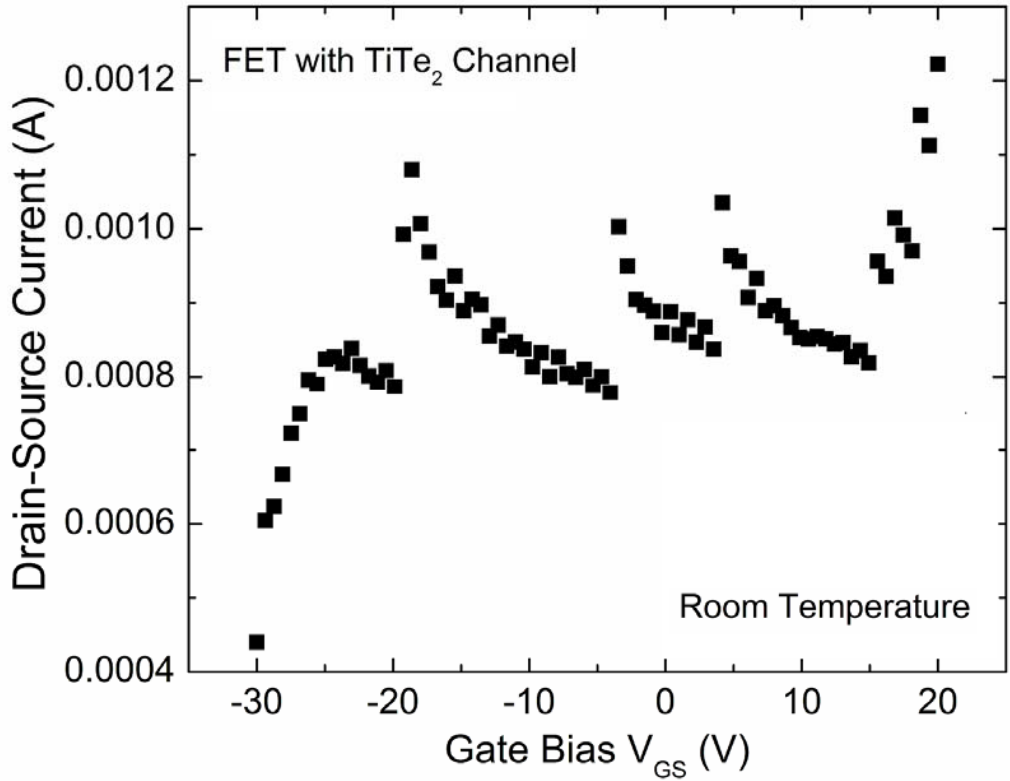

**Figure 4:** Drain current as the function of the back-gate voltage at $V_{DS}$=100 mV. Note a number of nearly equidistant peaks in I-V characteristics of the TiTe$_2$ FET at room temperature.

In a simple one-dimensional model, the wavevector of the CDW is $Q = 2k_F$, where $k_F$ is the Fermi wavevector. If periodic features in the $I$-$V_{GS}$ curve are related to the periodic modulation of a CDW phase, it is interesting to determine whether the change in the accumulated phase between the two contacts is approximately $2\pi$, i.e. is $\delta QL \approx 2\pi$? To estimate $\delta Q$, we must find $k_F$ in the $k_x - k_y$ plane. The electron density of 1.12 x 10$^{21}$ cm$^{-3}$ gives a sheet density per trilayer of $n_L$ = 7.28 x 10$^{13}$ cm$^{-2}$. The Fermi surface of TiTe$_2$ is complex with multiply disconnected parts in the $k_x - k_y$ plane. It has small dependence on $k_z$. An order of magnitude estimate of $k_F$ in the $k_x - k_y$ plane is $k_F^2 = 2\pi n_L$, and the change in $k_F$ is $\delta k_F = \pi \delta n_L / k_F$. The change in charge is $\delta n_L = \dfrac{C\delta V_{GS}}{qN}$ where $N$ = 154 is the number of trilayers in a 100 nm thick film and $C \approx 1.15 \times 10^{-8}$ F/cm$^2$ is the 300-nm back-gate oxide capacitance. Using $\delta V_{GS}$ = 10 V and $L$ = 8 µm, $\delta QL = 2\delta k_F L$ = 1.1 which is the same order of magnitude as $2\pi$.

The high sensitivity of the channel current to the gate voltage is difficult to explain considering that the screening length $\lambda_D = (\varepsilon k_B T / nq^2)^{1/2} = 0.36$ Å using $\varepsilon$=1 and $T$=300 K. A back-gate voltage should not be able to modulate the charge in a 100-nm thick sample. However, this same contradiction was apparent in the first demonstration of a CDW FET [27]. The screening length was one to two orders of magnitude smaller than the channel thickness, yet the channel current was still highly sensitive to the back-gate voltage. A number of different modulation mechanisms were considered, however, it was concluded that none of the mechanisms could easily account for the high sensitivity of the current to the back-gate voltage [27]. Prior studies of photoconduction in CDW materials [28-31] also revealed varying results: from to an order of magnitude increase in the linear current under the light illumination [28] to the absence of photoconduction [29-30].

While the exact physical mechanism of the observed multiple peaks in the drain current is not known one can envision a possible use of such I-Vs for information processing. Currently, there is a strong interest to alternative state variables – other than charges or electric currents – which can be used for information processing [32-35]. The abundance of non-linear physical phenomena observed in our devices provides possibilities for information processing. The intriguing I-V features, possibly related to the pinning - depinning CDW transitions, can be used in FET-type devices for implementing Boolean or non-Boolean logic gates. The design of such metal-semimetal logic gates can utilize the principles similar to those envisioned for the spin-wave devices [36-38]. The use of phases in CDW devices in addition to amplitudes is a powerful tool, which allows one to construct logic circuits with a fewer number of elements than required for CMOS technology.

In conclusion, we used the "graphene-like" exfoliation to prepare thin films of $TiTe_2$. The films were utilized as the channel layers in FET-type devices with Ti/Al/Au-metal contacts. Our measurements demonstrated intriguing non-linear I-Vs of the $TiTe_2$ FETs, which resemble those of CDW materials. Although no accurate explanation of the observed phenomena can be given at the present moment, we argue that the measured non-monotonic multi-peak characteristics can be potentially useful for the low-power information processing.

*Acknowledgments*

The work in Balandin and Lake groups was funded by the National Science Foundation (NSF) and Semiconductor Research Corporation (SRC) Nanoelectronic Research Initiative (NRI) project "Charge-Density-Wave Computational Fabric: New State Variables and Alternative Material Implementation" NSF-1124733 as a part of the Nanoelectronics for 2020 and Beyond (NEB-2020) program.